# Applying Science Models for Search

*Philipp Mayr, Peter Mutschke, Vivien Petras,
Philipp Schaer, York Sure*

**Abstract**

The paper proposes three different kinds of science models as value-added services that are integrated in the retrieval process to enhance retrieval quality. The paper discusses the approaches Search Term Recommendation, Bradfordizing and Author Centrality on a general level and addresses implementation issues of the models within a real-life retrieval environment.

## Introduction

Scholarly information systems often show three major points of failures, as pointed out in various studies (Mayr et al. 2008): (1) the vagueness between search and indexing terms, (2) the information overload by the amount of result records listed, and (3) the problem that pure term text based rankings, such as *tf-idf*, often do not meet the users' information need. Moreover, retrieval evaluations such as TREC and CLEF have shown that simple text-based retrieval methods scale up very well but do not progress anymore in terms of significant relevance improvements (Fuhr 2010, Armstrong et al. 2009).

The goal of the IRM[1] project (Mayr et al. 2008) therefore is to improve retrieval quality by value-added services that are based on computational models of the science system under study. The overall approach of IRM is to use models focusing on non-textual attributes of the research field, the scientific community respectively, as enhanced search stratagems (Bates 1990) within a scholarly information retrieval (IR) environment. This strongly meets the suggestion of Fuhr (2010) to move towards a more science model driven approach in IR which would lead to a broader view, an understanding of limitations of current models, and therefore the ability to open up alternative access paths into a field (Ingwersen & Järvelin 2005). The paper discusses the concepts of models on a general level and addresses implementation issues of the models within a real-life retrieval environment.

---
[1] http://www.gesis.org/irm/

## Model Discussion

Science models usually address issues in statistical modeling and visualization[2]. As a further dimension, that should be considered in science modeling as well, the paper focuses on the application of science models in IR (Mayr et al. 2011). Supposing that searching in a scholarly information system can be seen as a particular way of interacting with the science system, the overall assumption of our approach is that a user's search should improve by using science model driven search tactics. This approach meets the fact that the frequency of many structural attributes of the science system (such as co-authorships) usually follows some type of power-low distribution. These highly frequent attributes which are produced when applying the science models have a strong selectivity in the document space which can be utilized for IR.

The paper proposes three different kinds of science models as value-added services that are integrated in the retrieval process to enhance retrieval quality (see Figure 1): (1) a co-word analysis model for search term recommendations (STR), (2) a bibliometric model of re-ranking, called Bradfordizing, determining core journals for a field (BRAD), and (3) a network model of re-ranking examining the centrality of authors in scientific community (AUTH). STR addresses the problem of the vagueness between search and indexing terms, BRAD and AUTH the problem of large and unstructured result sets. In the following the models are discussed on a general conceptual level.

---

[2] See e.g. the workshop "Modelling Science" <http://modelling-science.simshelf.virtualknowledgestudio.nl/> and a forthcoming Special Issue in Scientometrics.

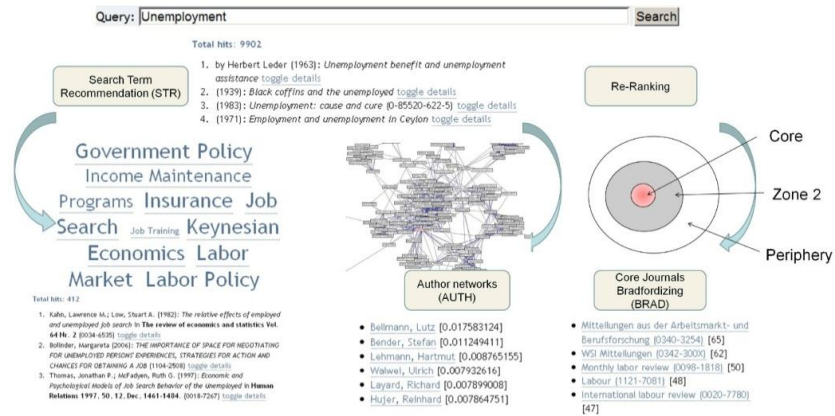

Figure 1: A simple search example (query term: "Unemployment") and typical structural attributes/outputs of implemented science models in our retrieval system. From left: Search Term Recommendation (STR) producing highly associated indexing terms, Author Networks (AUTH) with centrality-ranked author names and Bradfordizing based on Core Journals (BRAD) with highly frequent journal names/ISSNs.

## A Co-Word-Analysis Model for Query Expansion

Metadata-driven Digital Libraries share a common problem which Furnas (1987) and later Petras (2006) addressed as the "Language Problem in IR". Whenever a query is formalized the searcher has to come up with the "right" terms to best match the terms used in the index. Two languages domains have to match: (1) the language of scientific discourse which is used by the scientists who formulate the queries and (2) the language of documentation which is used by the database vendors. To overcome this query formulation problem and to provide a direct mapping between the language of discourse and the language of documentation Petras (2006) proposed a so called Search Term Recommender (STR). These recommenders are based on statistical co-word analysis and build associations between free terms (i.e. from title or abstract) and controlled terms (i.e. from a thesaurus). Controlled terms are assigned to the document during an intellectual or automatic indexation and enrich the available metadata on the document. The co-word analysis implies a semantic association between the free terms as instances of the language of discourse and the controlled terms as instances of the language of documentation. The more often terms co-occur in the text the more likely it is that they share a semantic relation. So, the model proposed focuses on the relationships among the terminological concepts describing the scientific discourse within a research field.

These semantic relations can be used to implement a query expansion mechanism where the initial query is expanded with a number of related controlled terms. Different evaluations of the Search Term Recommender as an approach for query expansion have shown (Petras 2005, Schaer et al. 2010) that co-word analysis based term suggestions significantly improve the precision of the retrieval process. Additionally they can provide an overview over different areas of discussion, which deal with particular concepts (perhaps assuming different meanings or directions of thought) when presented as an interaction method – for example in the form of a term cloud or a confidence ranked list.

This is especially true when domain-specific STR modules are used. A STR trained with a social science related document set will propose different terms and therefore concepts than e.g. a STR trained with documents from the domain of sport science. We may think of an query on "financial crisis": While the social science module will suggest terms like "stock market", "economic problems" of "international economic organizations" the other recommender will come up with relations to "sport economy", "player transfer" and "influence on performance". Each academic field has its own languages of discourse and documentation, so therefore the query suggestion methods have to adapt theses languages. The assumption is that term suggestions from several fields of research or information resources can provide a new view or different domain perspective on a topic (mainly in the interactive application of STRs). When used as an automatic query expansion mechanism this can lead to a phenomenon named "query drifts" (Mitra et al. 1998, Zighelnic & Kurland 2008) where the query and therefore the result set is transformed in a way the user didn't intend.

Beside query drifting, expanded queries tend to generate very large result sets (Efthimiadis 1996). Nevertheless in combination with a normal *tf-idf* ranking model positive effects which are related to the general concept of relevancy-ranking (Manning 2004) can be seen. By ranking the occurrences of both the user entered words and suggested terms from the STR, documents with a higher frequency are much more likely to be ranked in a top position. By expanding the query the result set automatically gets bigger (by OR-ing new terms) and at the same time the first hits are "narrowed down". This contradiction can be explained with the significantly higher discriminating power of the added terms and concepts in comparison to the terms of the original query which especially influences the term frequency part of the *tf-idf* formula.

## A Bibliometric Re-Ranking Model

For the problem of oversized result sets we propose a re-ranking model that applies a bibliometric law called Bradford law. Modeling science based on Bradford law is motivated by the necessity for researchers to concentrate on a small fraction of topically relevant literature output in a field. Fundamentally, Bradford law states that literature on any scientific field or subject-specific topic scatters in a typical way. In the literature we can find different names for this type of distribution, e.g. "long tail distribution", "extremely skewed", "law of the vital few" or "power law" which all show the same properties of a self-similar distribution. A Bradford distribution typically generates a core or nucleus with the highest concentration of papers – normally situated in a set of few so-called core journals – which is followed by zones with loose concentrations of paper frequencies. The last zone covers the so-called peripheral journals which are located in the model far distant from the core subject and normally contribute just one or two topically relevant papers.

Bradfordizing, originally described by White (1981), is a utilization of the Bradford law of scattering model which sorts/re-ranks a result set accordingly to the rank a scientific journal gets in a Bradford distribution. The journals in a search result are ranked by the frequency of their listing in the result set (number of articles in a certain journal). Bradfordizing assures that the central publication sources for any query are sorted to the top positions of the result set (Mayr 2009).

On an abstract level, re-ranking by Bradfordizing can be used as a compensation mechanism for enlarged search spaces with interdisciplinary document sets. Bradfordizing can be used in favor of its structuring and filtering facility. Our analyses show that the hierarchy of the result set after Bradfordizing is a completely different one compared to the original ranking. Furthermore, Bradfordizing can be a helpful information service to positively influence the search process, especially for searchers who are new on a research topic and don't know the main publication sources in a research field. The opening up of new access paths and possibilities to explore document spaces can be a very valuable facility. Additionally, re-ranking via bradfordized documents sets offer an opportunity to switch between term-based search and the search mode browsing. It is clear that the approach will be provided as an alternative ranking option, as one additional way or stratagem to access topical documents (cf. Bates 1990).

Interesting in this context is a statement by Bradford where he explains the utility of the typical three zones. The core and zone 2 journals are in his words "obviously and a priori relevant to the subjects", whereas the last zone (zone 3) is a very "mixed" zone, with some relevant journals, but also journals of "very general scope" (Bradford 1934). Pontigo and Lancaster (1986) come to a slightly different conclusion of their qualitative study. They investigated that experts on a topic always find a certain significant amount of relevant items in the last zone. This is in agreement with quantitative analyses of relevance assessments in the Bradford zones (Mayr 2009). The study shows that the last zone covers significantly less often relevant documents than the core or zone 2. The highest precision can very constantly be found in the core.

To conclude, modeling science into a core and a periphery – the Bradford approach – always runs the risk and critic of disregarding important developments outside the core. Hjorland and Nicolaisen (2005) recently started a first exploration of possible side effects and biases of the Bradford methods. They criticized that Bradfordizing favors majority views and mainstream journals and ignores minority standpoints. This is a serious argument, because by definition, journals which publish few papers on specific topics have very little chance to get into the core of a more general topic.

## A Network Model of Re-Ranking

Author centrality is a network model approach of re-ranking taking the social structure of a scientific community into account. The approach is motivated by the perception of "science (as) a social institution where the production of scientific knowledge is embedded in collaborative networks of scientists" (He 2009). The increasing significance of collaboration in science correlates with an increasing impact of collaborative papers (Beaver 2004), due to the complexity of nowadays research issues that require more collaboration (cf. Jiang 2008).

Collaboration in science is mainly represented by co-authorships between two or more authors who write a publication together. Transferred to a whole community, co-authorships form a co-authorship network as a particular "prototype of a social network" (Yin et al. 2006) that reflects the overall collaboration structure of a community. As inequality of positions is a structural property in social networks in general, locating strategic positions in scientific collaboration structures becomes an important issue also in examining the relevance of authors for a field (cf. Jiang 2008, Lu and Feng 2009, Liu et al. 2005). This perception of collaboration in science corresponds directly with the idea of structural centrality (Freeman 1977). Many authors characterize collaboration in science in terms that match a concept of centrality widely used in social network analysis (Chen et al. 2009, Yin et al. 2006), namely the betweenness centrality measure which evaluates the degree to which a node is positioned *between* others on shortest paths and thus emphasizes the node's brokerage role in the network's information flow (Freeman 1977, cf. Mutschke 2010).

As collaboration inherently implies the share of knowledge, high betweenness authors can be therefore seen as "pivot points of knowledge flow in the network" (Yin et al. 2006) and, by bringing different authors together, as the driving forces of the community making processes itself. The general assumption of the proposed model therefore is that the authors' impact on a scientific field can be quantified by their betweenness in co-authorship networks (cf. Yan and Ding 2009) and is therefore taken as an index of the of their publications. In short, this is done as follows (Mutschke 1994, 2004): (1) A co-authorship network is calculated on-the-fly on the basis of the result set to a specific query. (2) For each individual author in the network the betweenness is computed. (3) Each publication in the result set is weighted by the highest betweenness value of its authors (yielding a relevance value for each publication in the result set). (4) The result set is then re-ranked in descending order by that relevance values of the publications such that publications of central authors appear on top of the ranking.

The adequacy of this approach was confirmed by a number of empirical studies that turned out a high correlation between betweenness and other structural attributes, such as citation counts (Yan and Ding 2009), program committee membership (Liu et al 2005) and centrality of author topics in keyword networks (Mutschke and Quan-Haase 2001). Moreover, several studies have demonstrated that re-rankings based on network analysis methods can improve retrieval performance significantly (Yaltaghian and Chignell 2002, Zhou et al. 2007). Accordingly, an evaluation of the proposed ranking model (see below) has shown a higher precision than the text-based ranking. But, more importantly, it turned out that it favors quite other relevant documents. Thus, the true benefit of such a network model based ranking approach is that it provides a quite different view to the document space than pure text-based rankings.

However, two particular problems also emerge from that model. One is the conceptual problem of author name ambiguity (homonymy, synonymy) in bibliographic databases. In particular the potential homonymy of names may misrepresent the true social structure of a scientific community. The other problem is the computation effort needed for calculating betweenness in large networks that may bother, in case of long computation times, the retrieval process and finally user acceptance.

## Evaluation Results

To evaluate the general feasibility and performance of the models we conducted a user assessment where 369,397 single documents from the SOLIS database on Social Science topics were evaluated by 73 information science students for 10 topics. The documents include title, abstract, controlled keywords etc. The assessment system was built on top of the IRM prototype. The three services were compared to a *tf-idf* ranked result set from the underlying Solr search engine. Since the assessments were conducted with students instead of domain experts, Fleiss' Kappa values were calculated to measure the degree of inter-rater agreement (Schaer et al. 2010). Since there is no general accepted threshold for Fleiss' Kappa (cp. Sim and Wright, 2005) a custom threshold of 0.40 was selected and the values for three topics were dropped. The average precision among the top

10 documents for each service was: AUTH: 61%, BRAD: 56%, SOLR 52% and STR: 64%.

A comparison of the intersection of the relevant top 10 documents between each pair of retrieval service shows that the result sets are nearly disjoint. 400 assessed documents (4 services * 10 per service * 10 topics) only had 36 intersections in total. AUTH and SOLR as well as AUTH and BRAD have just three relevant documents in common (for all 10 topics), and AUTH and STR have only five documents in common. BRAD and SOLR have six, and BRAD and STR have five relevant documents in common. The largest, but still low overlap is between SOLR and STR, which have 14 common documents. Thus, there is no or very little overlap between the sets of relevant top-ranked documents obtained from different rankings.

Two results can be clearly seen: (1) The measured precision values of the evaluated services are at least the same or slightly better than the *tf-idf* based SOLR baseline (based on the degree of data cleaning) and (2) the services returned clearly disjoint result sets emphasizing that the three services provide quite different views to the document space. This strongly suggests thinking about a combination of the different services.

## Model Combination

As a next step in the IRM project we are dealing with combinations of the three models in various ways: (1) by using one model output as a filter mechanism for further iterations, (2) by computing combined ranking scores.

The first combination method works in a similar way as faceted search approaches (Tunkelang 2009) where items returned by different search services are used to filter the result set. Accordingly, AUTH can be applied on the set of publications assigned to core journals determined by BRAD (see Figure 2). Our prototype allows every combination of the three services. Typically the more filter steps are taken, the smaller the result set gets.

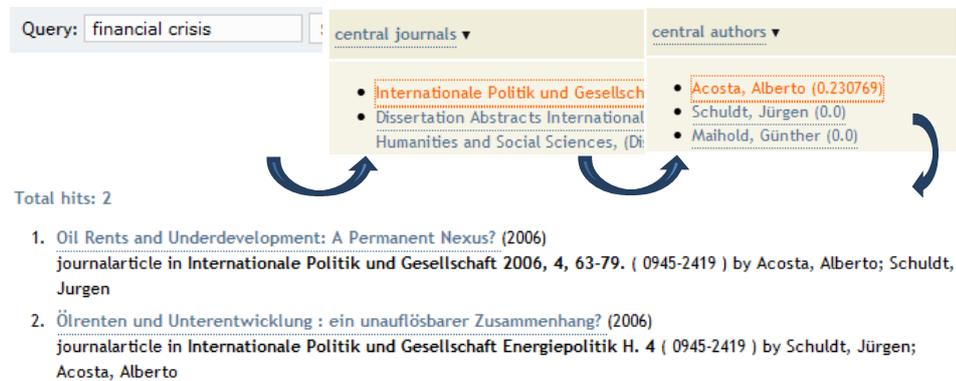

Figure 2: Filter workflow between BRAD and AUTH. Applying BRAD to the result set for 'financial crisis' yields the journal 'Internationale Politik und Gesellschaft' as the most relevant journal. Applying AUTH to the bradfordized result set yields Alberto as the most central author such that two articles of Acosta published in this journal.

A more sophisticated approach is to create a combined ranking score. As discussed before an inherent problem of both re-ranking mechanisms BRAD and AUTH is the lack of an "inner group" ranking. When a journal is detected as a core journal its corresponding documents are ranked to the top but the rank of each single document within this group is not defined. To solve this problem a combination of the original *tf-idf* score (mapped on [0,1]) and a journal or author specific weighting factor is applied.

To compute the weighting factor $W_j(q,d)$ for a document $d$ with respect to a journal $j$ and a query $q$ the document count for $j$ is multiplied with a factor of $1/J_{q,max}$ where $J_{q,max}$ is the maximum count for all journals $J$ obtained for $q$. This yields a score within [0,1]. The factor $W_j$ is 1 when $d$ is assigned to the journal having the highest coreness and it is 0 when $d$ is not published in a journal. The same approach is applied for the weighing factor for author centrality $W_a(q,d)$. Here all centrality values are mapped in [0,1] by multiplying each centrality value with $1/A_{q,max}$ where $A_{q,max}$ is the highest centrality value $q$. The factor $W_a$ is 1 when $d$ is assigned to most central author and it is 0 when $d$'s author is isolated.

The actual score, which is used for the final ranking process, is now computed with the following formula:

$$score(q,d) = tfidf(q,d) * W_j(q,d) * W_a(q,d),$$

where *tfidf* could be complemented by STR. When one of the factors is 0 the score is 0 and the document is discarded. Thus, the combined score tends to be a strong filtering method since it focuses on documents loading on all relevance indicators used.

## Outlook: A Service-Oriented Architecture of Retrieval Models

The proposed models are implemented in an interactive web-based prototype[3] using Solr for searching, Recommind Mindserver for the STR, own Java classes for BRAD and AUTH and the Grails Web framework for the interface. The user can dynamically modify the retrieval process by applying one of the models proposed either for the initial search or on the result set obtained. Moreover, the services can be combined to enhance the effects provided.

Currently, the prototype is going to be re-implemented as a service-oriented architecture (SOA) of re-usable, combinable and scalable web services. The major goal here is to have an architecture that provides services not only within the boundaries of a single IR system (as Private Services) but also as Public Services via the web such that the services can be used also by external information systems (see Figure 3). The other way around, this architecture allows for an easier integration of further value-added services provided by external partners.

---

[3] http://www.gesis.org/beta/prototypen/irm

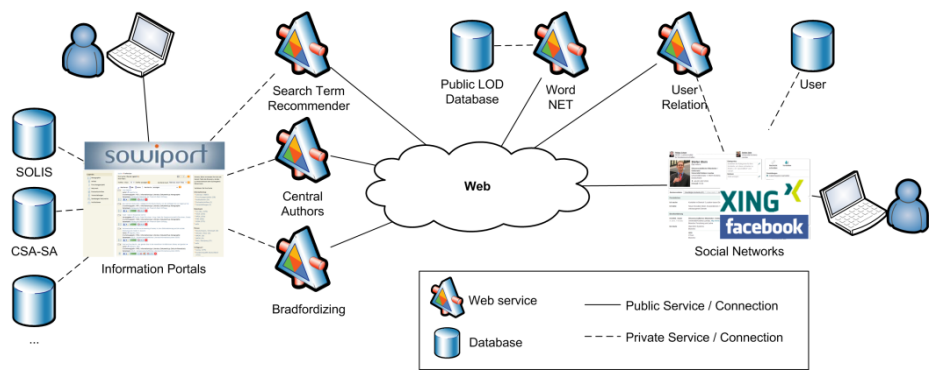

Figure 3: Retrieval services as loosely coupled Web Services in a service-oriented architecture. The three proposed services are used internally as private services. They are also available as public services on the web and are free to be integrated in other retrieval systems. At the same time external services e.g. from social networks or public services like Word Net can be integrated in our own system.

In this paper we have shown a further dimension of using science models, the application of science models for search. We have discussed and implemented three science model driven retrieval services to improve search in scholarly information systems. As a next step, our proposed SOA architecture might be an appropriate open framework for an integration and combination of further science models. This approach might be also a novel paradigm for enhanced Information Retrieval.

## References

Armstrong, T. G.; Moffat, A.; Webber, W. & Zobel, J. (2009). Improvements that don't add up: ad-hoc retrieval results since 1998. In CIKM '09: Proceeding of the 18th ACM conference on Information and knowledge management, pp. 601-610.

Bates, J. (1990), 'Where should the person stop and the information search interface start?', Inf. Process. Manage. 26 (5) , pp. 575-591.

Beaver, D. (2004): Does collaborative research have greater epistemic authority? Scientometrics 60 (3): 309-408.

Bradford, S. C. (1934). Sources of information on specific subjects. Engineering, 137(3550), 85-86.

Chen, C., Chen, Y., Horowitz, M., Hou, H., Liu, Z. and Pellegrino, D. (2009): Towards an explanatory and computational theory of scientific discovery. Journal of Informetrics 3: 191-209.

Freeman, L. C. (1977). A set of measures of centrality based on betweenness. Sociometry 40: 35-41.

Fuhr, N. (2010). IR Between Science and Engineering, and the Role of Experimentation. Keynote talk at CLEF 2010, Padua, Italy

Furnas, G. W.; Landauer, T. K.; Gomez, L. M. & Dumais, S. T. (1987). The Vocabulary Problem in Human-System Communication, Communications of the ACM 30 (11), pp. 964 – 971.

He, Z.-L. (2009): International collaboration does not have greater epistemic authority. JASIST 60(10): 2151-2164.

Hjørland, B., and Nicolaisen, J. (2005). Bradford's law of scattering: ambiguities in the concept of "subject". Paper presented at the 5th International Conference on Conceptions of Library and Information Science.

Ingwersen, P. and Järvelin, K. (2005), The turn: integration of information seeking and retrieval in context. Springer.

Jiang, Y. (2008): Locating active actors in the scientific collaboration communities based on interaction topology analysis. Scientometrics 74(3): 471-482.

Liu, X., Bollen, J., Nelson, M.L. and Sompel, H. van de (2005): Co-Authorship Networks in the Digital Library Research Community. Information Processing and Management 41 (2005): 1462–1480

Lu, H. and Feng, Y. (2009): A measure of authors' centrality in co-authorship networks based on the distribution of collaborative relationships. Scientometrics 81(2): 499-511.

Mayr, P.; Mutschke, P. & Petras V., (2008). Reducing semantic complexity in distributed digital libraries: Treatment of term vagueness and document re-ranking. In Library Review, 57 (3), pp. 213 – 224.

Mayr, P. (2009). Re-Ranking auf Basis von Bradfordizing für die verteilte Suche in Digitalen Bibliotheken. Humboldt-Universität zu Berlin

Mayr, P.; Mutschke, P., Schaer, P & Sure, Y. (2011 to appear). Science Models as Value-Added Services for Scholarly Information Systems. In Scientometrics.

Mutschke, P. (1994): Processing Scientific Networks in Bibliographic Databases. In: Bock, H.H., et al. (eds.): Information Systems and Data Analysis. Prospects-Foundations-Applications. Proceedings 17th Annual Conference of the GfKl 1993. Springer-Verlag, Heidelberg Berlin 127-133

Mutschke, P. and A. Quan-Haase, 2001: Collaboration and Cognitive Structures in Social Science Research Fields: Towards Socio-Cognitive Analysis in Information Systems. Scientometrics 52 (3): 487-502

Mutschke, P. (2004): Autorennetzwerke: Netzwerkanalyse als Mehrwertdienst für Informationssysteme. In: Bekavac, B. et al. (Hrsg.): Information zwischen Kultur und Marktwirtschaft: Proceedings ISI 2004. Konstanz, S. 141 – 162

Mutschke, P. (2010): Zentralitäts- und Prestigemaße. In: Häußling, Roger; Stegbauer, Christian (Eds.): Handbuch Netzwerkforschung. Wiesbaden: VS-Verlag für Sozialwissenschaften.

Petras, V. (2005). How one Word can make all the Difference - Using Subject Metadata for Automatic Query Expansion and Reformulation. Working Notes for the CLEF 2005 Workshop, 21-23 September.

Petras, V. (2006). Translating Dialects in Search: Mapping between Specialized Languages of Discourse and Documentary Languages. University of California, Berkley.

Pontigo, J. and Lancaster, F. W. (1986). Qualitative aspects of the Bradford distribution. Scientometrics, 9(1-2), 59-70.

Schaer, P.; Mayr, P. & Mutschke, P. (2010). Implications of Inter-Rater Agreement on a Student Information Retrieval Evaluation. In Proceedings of LWA2010 - Workshop-Woche: Lernen, Wissen & Adaptivitaet.

Sim, J. and Wright, C. C. (2005). The Kappa Statistic in Reliability Studies: Use, Interpretation, and Sample Size Requirements. In Physical Therapy. Vol. 85, pp. 257 – 268.

Tunkelang, D. (2009). Faceted Search, Morgan & Claypool Publishers.

Yaltaghian, B. and Chignell, M. (2002): Re-ranking search results using network analysis. A case study with Google. Proceedings of the 2002 conference of the Centre for Advanced Studies on Collaborative research.

Yan, E. and Ding, Y. (2009): Applying Centrality Measures to Impact Analysis: A Coauthorship Network Analysis. JASIST 60(10): 21-07-2118.

Yin, L., Kretschmer, H., Hannemann, R.A. and Liu, Z. (2006): Connection and stratification in research collaboration: An analysis of the COLLNET network. Information Processing & Management 42: 1599-1613.

White, H. D. (1981). 'Bradfordizing' search output: how it would help online users. Online Review, 5(1), 47-54.

Zhou, D., Orshansky, S.A., Zha, H. and Giles, C.L. (2007): Co-ranking authors and documents in a heterogeneous network. Proceedings of the 2007 Seventh IEEE International Conference on Data Mining: 739-744